\documentstyle[12pt,epsf]{article}
\newcommand{\CO}{{CompHEP}}
\newcommand{\noi}{\noindent}
\newcommand{\eeinto}   {$e^+e^- \longrightarrow \:$}
\newcommand{\geinto}   {$\gamma e \longrightarrow \:$}
\newcommand{\into}   {$\longrightarrow \:$}
\newcommand{\entb}{$e \nu t b \: $}
\newcommand{\hww}{$HWW \: $}
\newcommand{\nbbw}{$\nu \,b\, \bar{b}\, W\: $}
\newcommand{\wtb}{$W t b \: $}
\newcommand{\mt}{$m_{t} \: $}
\newcommand{\mh}{$M_{H} \: $}
\newcommand{\ee}{$e^+e^-\: $}
\newcommand{\vtb}{$|V_{tb}| \: $}

\newcommand{\bb}{$b\,\bar{b}\: $}

\newcommand{\qq}{$q\,\bar{q}\: $}

\newcommand{\gaga}{$\gamma\gamma\: $}

\newcommand{\znull}{$Z^0 \: $}
\newcommand{\hnull}{$H^0 \: $}
\newcommand{\gae}{$\gamma e\: $}

\newcommand{\SSQRTSGE}{$\sqrt{s_{\gamma e}}\:$}
\newcommand{\SSQRTSEE}{$\sqrt{s_{e^+e^-}}\:$}
\newcommand{\less}{\stackrel{ <}{\sim}}
\newcommand{\gess}{\stackrel{ >}{\sim}}

\def\PL #1 #2 #3 {Phys. Lett. {\bf#1}              (#3)  #2}
\def\NP #1 #2 #3 {Nucl. Phys. {\bf#1}              (#3)  #2}
\def\PR #1 #2 #3 {Phys. Rev. {\bf#1}               (#3)  #2}
\def\PP #1 #2 #3 {Phys. Rep. {\bf#1}               (#3)  #2}
\def\PRL #1 #2 #3 {Phys. Rev. Lett. {\bf#1}        (#3)  #2}
\def\CPC #1 #2 #3 {Comp. Phys. Commun. {\bf#1}     (#3)  #2}
\def\ANN #1 #2 #3 {Annals of Phys. {\bf#1}         (#3)  #2}
\def\APP #1 #2 #3 {Acta Phys. Pol. {\bf#1}         (#3)  #2}
\def\ZP  #1 #2 #3 {Z. Phys. {\bf#1}                (#3)  #2}
\def\NIM  #1 #2 #3 {Nucl. Instr. and Meth. {\bf#1} (#3)  #2}
 
\newcommand{\BS}{\bigskip}

\begin{document}
\pagestyle{empty}

\noi DESY 96-101

\BS\BS

\noi June 1996
\section*{
\vspace{4cm}
\begin{center}
\LARGE{\bf
 Higgs and Top Production in the Reaction
  \geinto \nbbw at TeV Linear Collider Energies
       }\\
\end{center}
}

\vspace{2.5cm}
\large
\begin{center}
E. Boos$^1$, A. Pukhov$^1$, M. Sachwitz$^2$ and H. J. Schreiber$^2$ \\ 
\bigskip \bigskip  
$^1$Institute of Nuclear Physics, Moscow State University, 119899,
Moscow, Russia \\
$^2$DESY-Institut f\"{u}r Hochenergiephysik, Zeuthen, FRG \\
\end{center}
\newpage
\pagestyle{plain}
\pagenumbering{arabic}
\large
\begin{abstract}
For an electron-photon collider the complete tree-level cross sections
of the reaction \geinto \nbbw \, are computed at center-of-mass
energies between 0.5 and 2.0 TeV, for top masses of 160 to 200 GeV and 
Higgs masses between 80 and 140 GeV within the Standard Model.
It is shown that most of the \nbbw \, events are due to Higgs and
$Z/\gamma^*$ production (with $H, Z/\gamma^*$ \into \bb \, decay) 
while top production (with $t$ \into $bW$ decay) is about 50\% smaller.
Multiperipheral background and interferences are small, respectively
negligible, in the energy range studied.
By convoluting the basic cross sections with an energy
spectrum of the backscattered photon beam, and inserting linear
collider luminosities as anticipated in present designs,
realistic \nbbw \, event rates are estimated.
This results in large event rates for \geinto $\nu t b$ and \geinto $\nu
H W$.
We estimate that the CKM matrix element \vtb \, can be probed from the
$\nu t b$ final state to an accuracy of 1-3\% at \SSQRTSEE $\gess$ 1
TeV.
Assuming an effective Lagrangian based on dimension-6 operators we
discuss the sensitivity for detecting deviations
of the \hww \, coupling from the Standard Model in the reaction
\geinto $\nu H W$.
\end{abstract}

\section{Introduction}

Electron-photon and photon-photon colliders  are
seriously considered as interesting options to upgrade future linear
\ee  colliders.
The electron-photon and  photon-photon collisions have been 
to a great extent studied in the
context of \ee  physics by using virtual
bremsstrahlung and beamstrahlung photons.
 However,  \gae \, and \gaga colliders 
 where the photon beams are generated by backward Compton
 scattering of laser light on the high energy electron beams
supplied by the underlying \ee \,  collider, have great advantages.
The mechanism of the electron to photon beam conversion has been
studied theoretically by many authors, see e.g. \cite{ginzburg,milburn}. 
According to their results, the typical photon beam energy is about
0.7 times the electron beam energy.
Thus, the \gae \, collision energy is roughly 80\% of the
underlying \ee \, collider energy.
If the polarizations of the source electron beam and the laser photon
can be controlled, the resulting photon beam can be almost
monochromatic and highly polarized. 

With increasing collision energy the complexity of the 
events is expected to increase. 
While at the \znull \, peak (LEP-1 energy) two-fermion final states
produced by the \znull \, decay are, by far, the most important
processes, 4-fermion final states become
dominant at LEP-2 due to $W^+W^-$ production.
It is expected that at higher energies many-body reactions with
fermions and vector bosons in the final state will play a growing
part. 
Such reactions may yield additional or complementary, or even more
stringent, information about couplings of e.g. the Standard Model (SM)
\cite{sm} 
Higgs boson and the top quark, or may reveal deviations from SM
predictions in more obvious ways.

In this paper we analyze the reaction 

\begin{equation}
\gamma \quad e \longrightarrow \nu \quad b \quad \bar{b} \quad W^-
\label{eq:main}
\end{equation}

\noi at high center of mass (cm) energies.
We do not focus on low energy \gae \, reactions from
Weizs\"{a}cker-Williams \cite{weizwilliams} and beamstrahlung photons
which will be automatically generated in \ee \, collisions.
We assume that the backscattered photon beam is
unpolarized and that, on the average, the number of the backscattered
photons produced per positron is close to 1. 
Interactions of quarks (and gluons), created by
perturbative \qq (gluons) fluctuations of the real photon, 
are not taken into account in the present paper.

Reaction (\ref{eq:main}) is  interesting on its own because it
simultaneously involves single top quark production in the subreaction 

\begin{equation}
\gamma \quad e \longrightarrow \nu \quad \bar{b} \quad t
\label{eq:2}
\end{equation}

\noi and associated Higgs boson production in

\begin{equation}
\gamma \quad e \longrightarrow \nu \quad W^- \quad H^0,
\label{eq:3}
\end{equation}

\noi with subsequent decays of the top, $t$ \into $W b$, and 
respectively, the Higgs, $H$ \into \bb.
Both reactions have been studied in the past \cite{boos1,jikia} 
and their abundant rates were emphasized.
We note that reaction (\ref{eq:2}), a subchannel of
the process \eeinto \entb \, recently studied by the authors \cite{boos2},
serves as a unique tool to probe the \wtb \, coupling and to measure
the Cabibbo-Kobayashi-Maskawa (CKM) marix element \vtb \, with high
precision. 
Reaction (\ref{eq:3}) will, due to its large rate, be of special interest
for probing the \hww coupling.

In this paper we present results of complete 
tree-level calculations of the
reaction \geinto \nbbw within the SM.
Decays of unstable particles with correct spin structures 
and contributions from all nonresonant diagrams 
are taken into account.
In this way, the subreactions (\ref{eq:2}) and (\ref{eq:3}) are involved
automatically in the 2-to-4 body calculations 
and, as will be shown, they can easily be extracted from
the inherent background leading to the same final state.

The paper is organized as follows.
In sect. \ref{sec2} we introduce the contributing diagrams, 
the method of
calculation, the problem of singularities and the choice of a proper
kinematical scheme.
Cross sections for reactions (\ref{eq:main}), (\ref{eq:2}) and
(\ref{eq:3}) are presented in sect. \ref{sec3}
as function of the photon-electron cm energy
\SSQRTSGE.
Higgs and top masses are varied between 80 and 140 GeV and 
160 and 200 GeV, respectively. 
In sect. \ref{sec4} the cross sections of reactions
(\ref{eq:main})-(\ref{eq:3})  are
folded with an assumed energy spectrum of the backscattered photon beam.
In this way, the impact of the model-dependent photon spectrum
of ref. \cite{ginzburg} on the
expected event rate is illustrated.
Sect. \ref{sec5} is devoted to the prospects of measuring the matrix
element \vtb, while in sect. \ref{sec6} the detection of possible anomalous
Higgs couplings to $W^+W^-$ bosons is discussed.
Sect. \ref{sec7} contains the summary.

\section{Diagrams, singularities and kinematical scheme}
\label{sec2}
   
\begin{figure*}[hbtp]
\mbox{\epsfxsize=17cm\epsfysize=19cm\epsffile{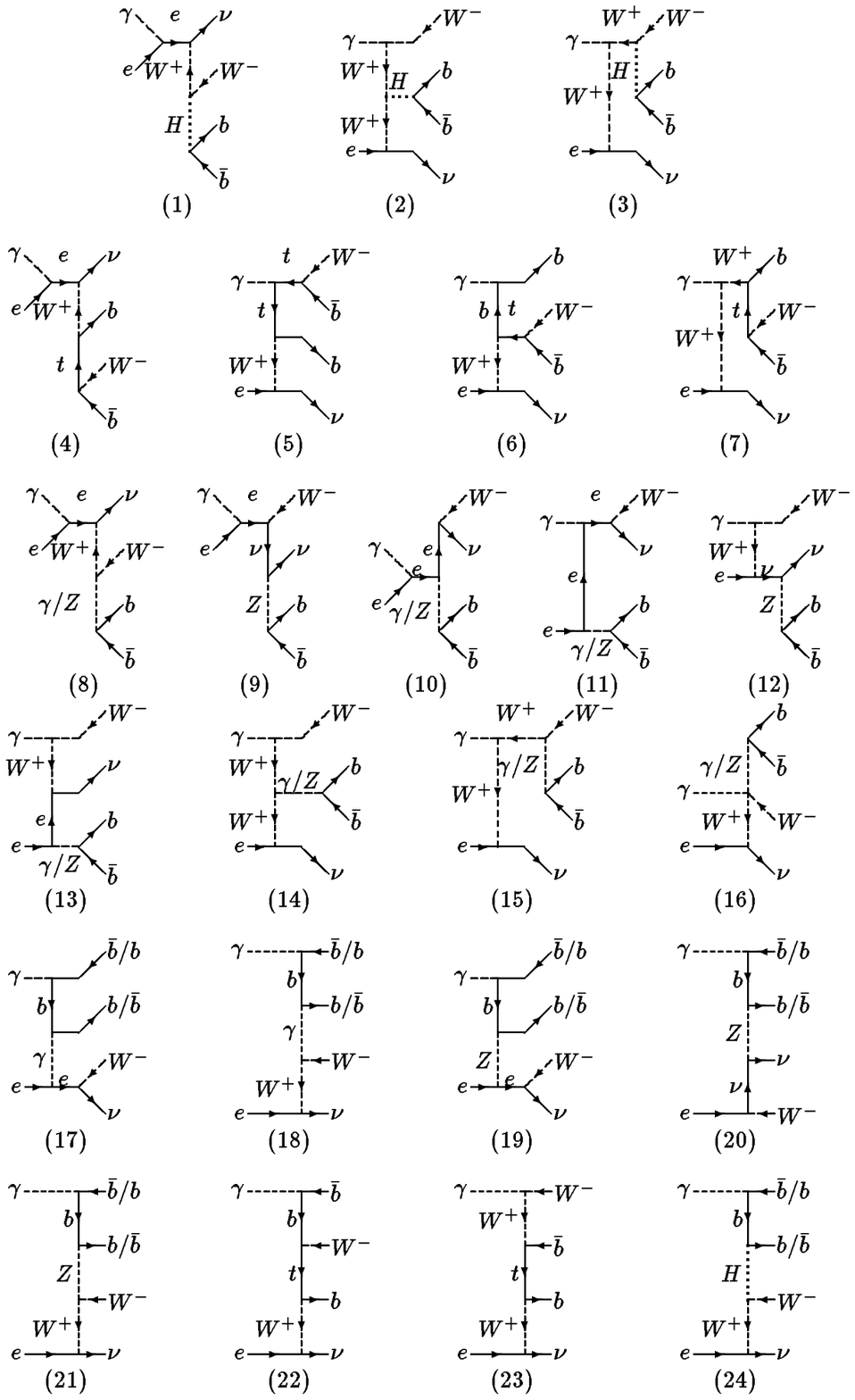}}
\caption{ Feynman diagrams for the reaction \geinto \nbbw.}
\label{fig:feyn}
\end{figure*}

All SM tree-level diagrams contributing to the reaction \geinto \nbbw
\, are shown in Fig. \ref{fig:feyn}.
Only contributions from the physical particles are presented.
Diagrams for Faddeev-Popov's ghosts and Goldstone bosons are omitted. 
Their contributions were however taken into account 
in the t'Hooft-Feynman gauge.

The three diagrams in the first row of Fig. \ref{fig:feyn} involve the
Higgs boson production with its decay to \bb.
Top production (with subsequent $t$ \into $b W$ decay) occurs in diagrams
(4)-(7) in the second row.
Both these classes of diagrams are denoted {\it 'signal diagrams'} in the
following. 
In diagrams (8)-(16) $Z$ and virtual photons are
produced with subsequent $Z/\gamma^*$ \into \bb \, decays.
The remaining diagrams of Fig. \ref{fig:feyn} are of multiperipheral
nature. 

The results presented have been obtained by means of the computer
package \CO \, \cite{comphep}.
The present version of \CO \, performs analytic calculations of the
matrix elements squared, generates an optimized Fortran code and
generates a flow of events.
In addition, it provides the possibility for the
user to choose an appropriate kinematical scheme.
  
The basic input parameters used in the program are taken  
from the report of the Particle Data Group
\cite{pdg} or are as listed here:
 $m_b$    =  4.3 GeV, $\alpha_{EW}$ =1/128,
\vtb = 0.999, $M_Z$    =  91.187 GeV,   
 $\sin^2\Theta_W$    =  0.23, $M_W=M_Z*\cos\Theta_W$, $\Gamma_Z$=2.50 GeV
 and $\Gamma_W$=2.09 GeV.

For unstable particles, Breit-Wigner formulae have been used for
the s-channel propagators.
For the Higgs and top the tree-level widths are applied.

As seen from the diagrams in Fig. \ref{fig:feyn}, a number of
singularities exists in the s- and t-channels.
In the phase space integration by the adaptive Monte Carlo method, a
proper treatment of such singular behaviour is necessary in order to
obtain stable results.
Usually, singularities are smoothed by appropriate transformations of
variables, and ref. \cite{pukhov} describes formulae for smoothing of
singularities as used in \CO, which have been adopted in our calculations.

Smoothing of singularities works effectively only in the case of a
proper choice of the kinematical scheme.
Basically, we selected integration variables in such a way 
that each of the singularities occurs in only one of these variables.
\begin{figure}[hbtp]
\mbox{\epsfxsize=17cm\epsfysize=5cm\epsffile{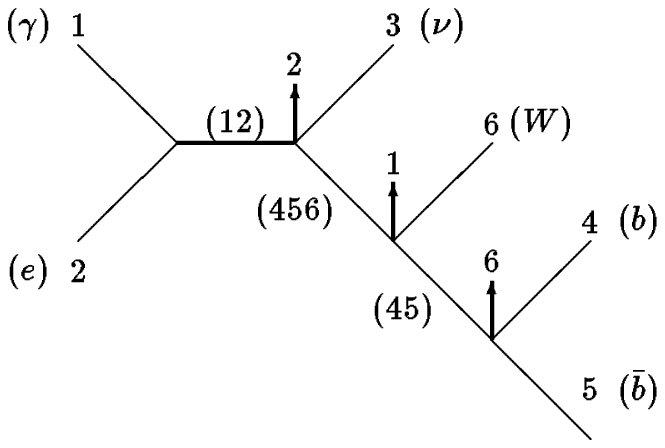}}
\caption{ Illustration of the kinematical scheme used in the
  calculation.}
\label{fig:kine}
\end{figure}
The choice of the scheme used for reaction (\ref{eq:main}), \geinto
\nbbw, is briefly described in the following, see also
Fig. \ref{fig:kine}: 
\begin{itemize}
\item in the first step we consider the decay of cluster (12) into
  particle (3) and cluster (456) in its rest frame.
  The angle between the three-momenta of particles(2) and (3),
  $\Theta_{23}$, whose cosine is a linear function of the
  momentum transfer-squared $t_{23} \equiv t_{e\nu}$, has been chosen
  as a variable, with a singularity at $t_{23} = M_W^2$;
\item next, the decay of cluster (456) into particle (6) and cluster
  (45), considered in its rest frame, offers two angles as variables:
  $\Theta_{16}$, the angle between particle (1) and (6), and
  $\Theta_{1(45)}$, that between particle (1) and cluster (45).
  These angles are linearly related to the momentum transfer-squared
  $t_{16} \equiv t_{\gamma W}$, and  $t_{1(45)} \equiv t_{\gamma
    (b\bar{b}})$, respectively, having singularities at $M_W^2$;
\item selecting the invariant mass of cluster (45), $M_{b\bar{b}}$, as
  a variable, singularities occur at its threshold and at $M_Z$ and
  \mh;
\item finally, the decay of cluster (45) into particles (4) and (5)
  considered in its rest frame, offers the angle $\Theta_{56}$ as a
  variable, which is linearly related to $M_{Wb}^2$, with a singularity
  at $m_{t}^2$.
\end{itemize}
The singularities mentioned above turned out to be the most important ones
and were smoothed according to the description of ref. \cite{pukhov}.
In this way we are confident in the stability of our Monte Carlo
results. 

\section{Basic cross section results}
\label{sec3}
 
In possible future \gae \, colliders, the photon beam will be produced
by backscattering of laser light on the high energy
electron beam.
The photon beam expected from such scatterings will have a nontrivial
energy-dependent luminosity spectrum.
Its measurement will be crucial so that 
convolutions with (theoretical) \gae \, reaction cross sections
can be performed with high
confidence at a given \ee \, cm energy \SSQRTSEE.
At present, such convolutions must be carried out by means of
model-dependent photon spectra.
Therefore we present in this paper at first the unconvoluted
cross sections for reaction (\ref{eq:main}) and updated cross sections
for reactions (\ref{eq:2}) and (\ref{eq:3}), as a function of the \gae \,
cm energy \SSQRTSGE.
Based on these results the reader can, by using his
preferred photon spectrum, carry out convolutions in order to obtain
'realistic' event rate expectations.

\begin{figure}[htbp]
\mbox{\epsfxsize=17cm\epsfysize=9cm\epsffile{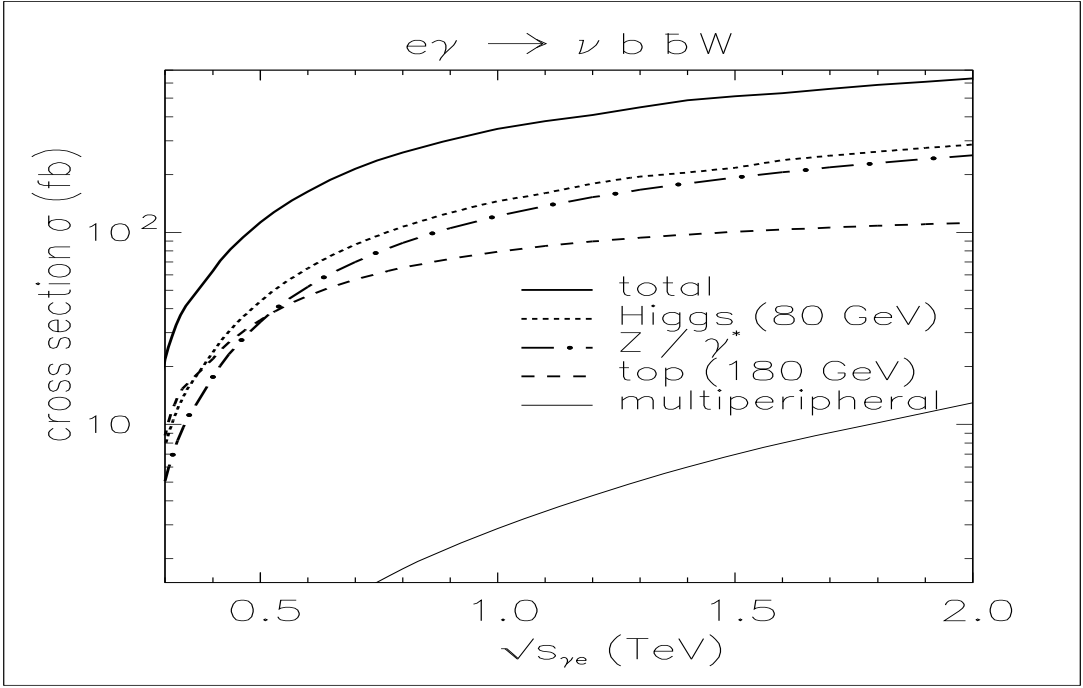}}
\caption{Total cross section for the reaction   \geinto \nbbw
as function of the $\gamma e$ cm energy,  
for \mt = 180 GeV and \mh = 80 GeV. Also the individual 
contributions of the Higgs, the top, the $Z/\gamma^*$ and the
multiperipheral diagrams are shown.}
\label{fig:2}
\end{figure}

Fig. \ref{fig:2} shows the total cross section for reaction
(\ref{eq:main}), \geinto \nbbw, as function of \SSQRTSGE, for a Higgs
mass of \mh = 80 GeV and a top mass of \mt = 180 GeV.
It rises with increasing energy over the whole energy
range considered.  
Also shown are the cross sections for the single top reaction
(\ref{eq:2}), the Higgs reaction (\ref{eq:3}), the
$Z/\gamma^*$ and the multiperipheral contributions corresponding to
the different classes of diagrams as discussed in sect. \ref{sec2}.
The Higgs and the $Z/\gamma^*$ rates are very close to each other.
Top production rises less strongly with increasing  energy 
such that in the 1-2 TeV region, it contributes only about half as
much as each of the other two channels.
Multiperipheral contributions are, as expected from previous 4-body
final state investigations \cite{boos2}, considerably weaker.
Still, they are to some extent also responsible for the continuous 
rise of the overall reaction cross section at high energies.

Table \ref{tab:1} summarizes the cross sections discussed at \SSQRTSGE =
0.5, 1.0, 1.5 and 2.0 TeV.

\BS\BS

\begin{table*}
\caption[ ]{Cross sections (in fb) for reaction (\ref{eq:main}) and
  different subchannels as discussed in the text as well as their
  interferences. } 
\begin{tabular}{p{4cm}lllll}    
 & &  & & \\
\hline\noalign{\smallskip}
 & &  & & \\
\SSQRTSGE, TeV   & 0.5        & 1.0 & 1.5 & 2.0  \\
 & &  & & \\
\hline\noalign{\smallskip}
 & &  & & \\
 $\sigma$(total)        & 113(1)\hspace{1.3cm}   
              & 346(4)  \hspace{1.3cm}
              & 511(5)  \hspace{1.3cm}
              & 634(8)   \hspace{1.3cm} \\
 & &  & & \\
 $\sigma$(\mh=80 GeV)     & 43.7(1)   
              & 145.2(2)   
              & 224.2(3)  
              & 286.3(4) \\
 & &  & & \\
 $\sigma$(\mt=180 GeV)     & 34.9(1)   
              & 79.3(1)   
              & 100.6(2)  
              & 112.2(2) \\
 & &  & & \\
 $\sigma$($Z/\gamma^*$)     & 34.2(1)   
              & 121.7(4)   
              & 193.0(7)  
              & 252.0(9) \\
 & &  & & \\
 $\sigma$(Multiperipheral)     & .588(2)   
              & 2.87(2)   
              & 6.91(8)  
              & 13.0(2) \\
 & &  & & \\
 Interferences     & -1.01(2)   
              & -4.2(1)   
              & -11.7(2)  
              & -23.7(3) \\
 & &  & & \\
\hline\noalign{\smallskip}
\end{tabular}
\label{tab:1}
\end{table*}
The total rates given in the first row are accurate to  $\less$ 
1.5\% by virtue of the
choice of an adequate kinematical scheme, despite of non-vanishing masses of
the participating particles.
All diagrams of Fig. \ref{fig:feyn} have been  included in the
calculations.
The other numbers in Table \ref{tab:1} obtained from 
subsets of diagrams 
have a somewhat better accuracy of $\less$ 0.2\%.
The interferences between
different classes of diagrams shown in the last row 
rise with increasing energy to significant non-zero values at
\SSQRTSGE = 1.5-2 TeV.
Their behaviour is similar, apart from the sign, to that
of the multiperipheral cross section so that the two contributions cancel
 each other to some extent. 
Since interferences between the signal and the $Z/\gamma^*$ diagrams 
are consistent
with zero, the cross section of reaction (\ref{eq:main}) is 
well approximated  by their incoherent sum.

\begin{figure}[htbp]
\mbox{\epsfxsize=17cm\epsfysize=9cm\epsffile{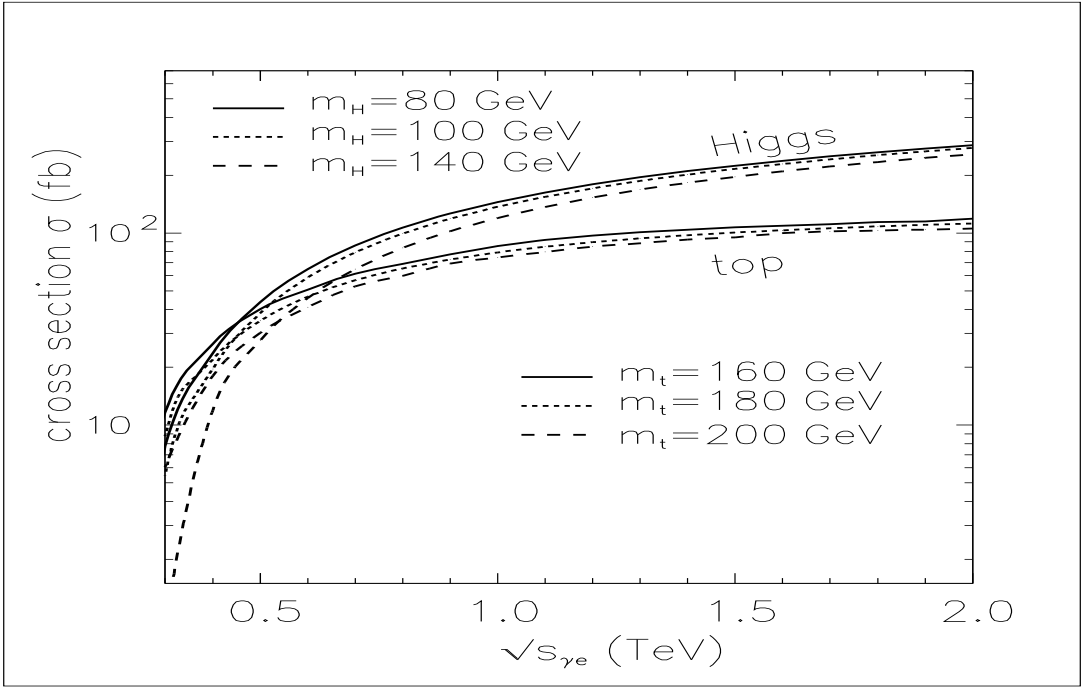}}
\caption{Cross sections for the top and Higgs subreactions as function
of the $\gamma e$ cm energy,  
for $m_{top}$ = 160, 180 and 200 GeV and \mh = 80, 100 and
140 GeV.}
\label{fig:3}
\end{figure}

The dependence of the cross section for reaction (\ref{eq:2}) on the
top mass \mt \, and that for reaction (\ref{eq:3}) on the Higgs mass,
\mh, is illustrated in Fig. \ref{fig:3}, again as a function of
\SSQRTSGE.
\mh \, is varied between 80 and 140 GeV, while \mt is assumed to be 
 160, 180 and 200 GeV.
Away from the thresholds neither a strong Higgs mass nor a strong top mass
dependence is observed.
The cross sections in each case decrease with increasing 
mass of the particle in question, as expected from
phase space.

It is obvious from Figs.\ref{fig:2} and \ref{fig:3} and Table
\ref{tab:1} that at large \SSQRTSGE \, and moderate Higgs
masses (80-140 GeV) the cross sections for reactions
(\ref{eq:main}) - (\ref{eq:3}) are large so that they become very
interesting, from an experimental point of view, for more detailed
studies.

\section{Backscattered photon spectrum convoluted cross sections}
\label{sec4}

In order to obtain 'realistic' estimations of event
rates expected for reactions (\ref{eq:main})-(\ref{eq:3}) the cross
sections of sect. \ref{sec2} have to be convoluted with the backscattered
photon flux.
We have, as an example, adopted in our calculations the photon spectrum
as proposed in ref. \cite{ginzburg}\footnote{Here the source electron beam and
  the laser photon beam are assumed to be unpolarized.}

\begin{displaymath}
F_{\gamma}  =  \frac{1}{N(x_0)}\left[1 - y + \frac{1}{1-y} -
\frac{4y}{x_0(1-y)} + \frac{4y^2}{x_0^2(1-y)^2} \right]   \nonumber 
\end{displaymath}
\noi with
\begin{eqnarray} 
N(x_0) & = & \frac{16 + 32x_0 + 18x_0^2 + x_0^3}{2x_0(1+x_0)^2} \quad +
\nonumber \\
[1cm]  
& & +  \quad \frac{x_0^2 - 4x_0 - 8}{x_0^2}\ln (1+x_0) ,
\label{eq:laser}
\end{eqnarray}

where the parameter $x_0$ depends on the laser photon frequency.
It should be chosen such that possible onset of \ee \, pair
production between laser photons and backscattered photons is avoided.
This constraint leads to an $\gamma e$ energy spectrum which peaks
close to its maximum at 0.83 \SSQRTSEE.
'Realistic' cross sections are then evaluated by folding the
basic SM cross section of sect. \ref{sec2} with the photon spectrum
(\ref{eq:laser}). 
The results so obtained are summarized in Tab. \ref{tab:3}, for the total, the
Higgs (\mh = 80 GeV), the top (\mt = 180 GeV) and the $Z/\gamma^*$ 
contributions at \SSQRTSEE = 0.5, 1.0, 1.5 and 2.0 TeV.

\begin{table*}[htbp]\centering
\caption[ ]{Laser spectrum convoluted cross sections (in fb) of reaction
  (\ref{eq:main}) and different subreactions as discussed in the text
  at different \ee \, cm energies.} 
\begin{tabular}{p{4cm}lllll}    
 & &  & & \\
\hline\noalign{\smallskip}
 & &  & & \\
\SSQRTSEE, TeV   & 0.5        & 1.0 & 1.5 & 2.0  \\
 & &  & & \\
\hline\noalign{\smallskip}
 & &  & & \\
 $\sigma$(total)        & 41.5(7)\hspace{1.3cm}   
              & 187(2)  \hspace{1.3cm}
              & 314(8)  \hspace{1.3cm}
              & 420(4)   \hspace{1.3cm} \\
 & &  & & \\
 $\sigma$(\mh=80 GeV)     & 15.6(1)   
              & 76.8(1)   
              & 136.3(2)  
              & 186.9(2) \\
 & &  & & \\
 $\sigma$(\mt=180 GeV)    & 14.4(1)   
              & 49.0(1)   
              & 71.8(1)  
              & 86.7(2) \\
 & &  & & \\
 $\sigma$($Z/\gamma^*$)   & 11.4(1)   
              & 62.1(4)   
              & 113(1)  
              & 159(1) \\
 & &  & & \\
\hline\noalign{\smallskip}
\end{tabular}
\label{tab:3}
\end{table*}

Clearly, the cross section reduction due to the convolution 
is largest at
0.5 TeV (about 50\%) and becomes smaller with increasing energy;
e.g. at 2 TeV an event loss of $\sim$ 30\% is expected.
For completeness, Tabs. \ref{tab:4} and \ref{tab:5} show the Higgs and
top cross sections for various \mh and \mt \, values as a
function of \SSQRTSEE.

\begin{table*}[htbp]\centering
\caption[ ]{Laser spectrum convoluted cross sections (in fb) of reaction
  (\ref{eq:3}) for different \ee \, cm energies and Higgs masses.} 
\begin{tabular}{p{4cm}lllll}    
 & &  & & \\
\hline\noalign{\smallskip}
 & &  & & \\
\SSQRTSEE, TeV   & 0.5        & 1.0 & 1.5 & 2.0  \\
 & &  & & \\
\hline\noalign{\smallskip}
 & &  & & \\
$\sigma$(\mh=80 GeV)     & 15.6(1) \hspace{1.3cm}   
              & 76.8(1)    \hspace{1.3cm}
              & 136.3(2)   \hspace{1.1cm}
              & 186.9(2)  \hspace{1.1cm} \\
 & &  & & \\
$\sigma$(\mh=100 GeV)    & 12.8(1)   
              & 70.7(1)   
              & 128.8(2)  
              & 178.7(3) \\
 & &  & & \\
$\sigma$(\mh=120 GeV)    & 10.2(1)   
              & 64.5(1)   
              & 121.0(2)  
              & 170.2(3) \\
 & &  & & \\
$\sigma$(\mh=140 GeV)    & 7.81(1)   
              & 58.4(1)   
              & 113.0(2)  
              & 161.1(3) \\
 & &  & & \\
\hline\noalign{\smallskip}
\end{tabular}
\label{tab:4}
\end{table*}

\begin{table*}[htbp]\centering
\caption[ ]{Laser spectrum convoluted cross sections (in fb) of reaction
  (\ref{eq:2}) for different \ee \, cm energies and  top masses.} 
\begin{tabular}{p{4cm}lllll}    
 & &  & & \\
\hline\noalign{\smallskip}
 & &  & & \\
\SSQRTSEE, TeV   & 0.5        & 1.0 & 1.5 & 2.0  \\
 & &  & & \\
\hline\noalign{\smallskip}
 & &  & & \\
$\sigma$(\mt=160 GeV)    & 17.5(1)   \hspace{1.3cm}
              & 54.0(1)   \hspace{1.3cm}
              & 77.1(2)  \hspace{1.3cm}
              & 91.8(2)   \hspace{1.3cm} \\
 & &  & & \\
$\sigma$(\mt=180 GeV)    & 14.4(1)   
              & 49.0(1)   
              & 71.8(1)  
              & 86.7(2) \\
 & &  & & \\
$\sigma$(\mt=200 GeV)    & 11.8(1)   
              & 45.1(1)   
              & 67.9(1)  
              & 82.8(2) \\
 & &  & & \\
\hline\noalign{\smallskip}
\end{tabular}
\label{tab:5}
\end{table*}

It is encouraging that 
even after degradation of the basic cross sections by 
appropriate photon flux convolution, an electron-photon
 collider can considerably improve the
physical capabilities of Higgs and top studies.
Examples are presented in the next section.
It is also worthwhile to note that  event rates
for reactions (\ref{eq:main})-(\ref{eq:3}) from Weizs\"{a}cker-Williams and
beamstrahlung photons in an underlying \ee \, collider are 
expected to be significantly 
below the laser induced \gae \, collision rates \cite{boos1,boos2}.

\section{$W t b$ coupling and the measurement of the matrix element \vtb}
\label{sec5}

Measurements of \vtb \, or the partial width $\Gamma_{tWb}$, which are
related in the SM, are known to be nontrivial.
Recently it has been suggested to use the single top quark reaction 
\eeinto \entb \, at high energies \cite{boos2} which offers the possibility to
obtain a relatively precise value of \vtb.
In this channel, the \vtb \, measurement capability relies mainly on
the Weizs\"{a}cker-Williams photon exchange contributions, $\gamma^*
e$ \into $\nu b t$. 
Using however 
the laser backscattered high energy photon beam instead of $\gamma^*$ the
cross sections are typically enhanced by a factor of 3-5. 
As outlined in the previous sections, reaction (\ref{eq:main})
involves to a great extent $\nu b t$  events involving
the $W t b$ coupling.
Their rate is directly proportional to
\vtb$^2$.

\begin{figure}[htbp]
\mbox{\epsfxsize=17cm\epsfysize=9cm\epsffile{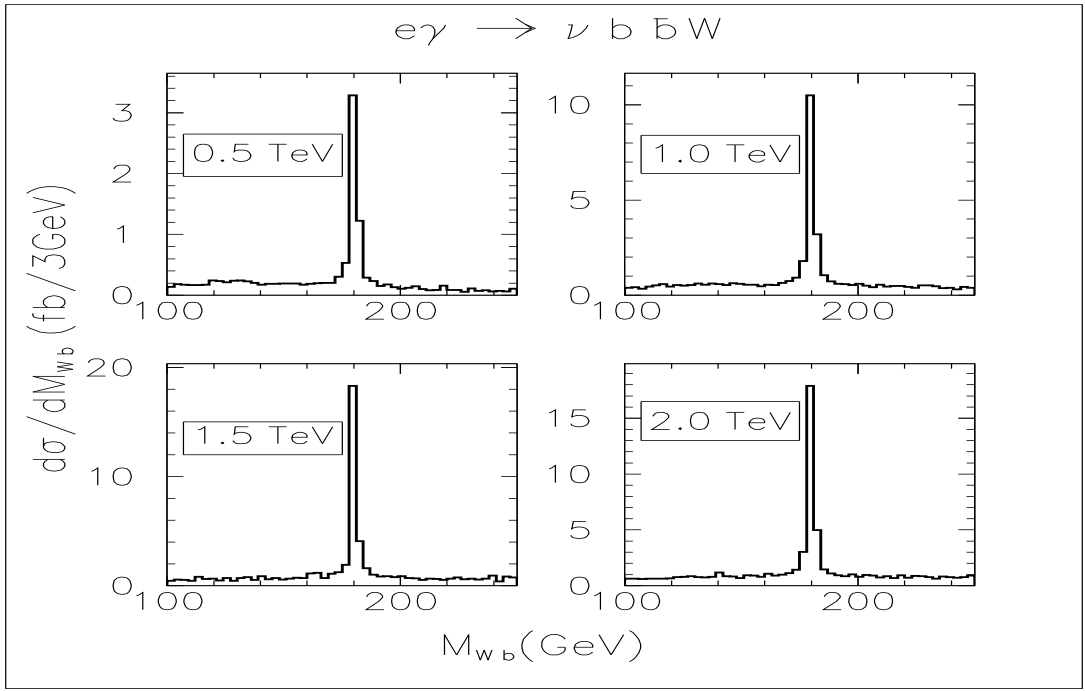}}
\caption{Differential cross sections d$\sigma$/d$M_{Wb}$ 
  of reaction (1) as
  function of $M_{Wb}$.}
\label{fig:4}
\end{figure}

The top quark is not observed directly; it decays into a $W$ and a
$b$ quark, leading to the final state \nbbw.
The extraction of the top out of this 4-body final
state can however be easily achieved as seen 
from Fig. \ref{fig:4}, where the invariant masses
of the $W$ and the $b$ are shown for four energies.
Clear top signals exist on a small and smooth background.
Selection of the top event requires only a cut around \mt \, and no
further demands.

Using expected \ee \, luminosities as proposed in ref. \cite{Wiik-TESLA}, an
$e-\gamma$ conversion factor of 0.8 and a 30\% $\nu b t$ event
detection probability (due to cuts to observe the top decay products
and the $b$-jet and to eliminate backgrounds; major backgrounds are
expected from the reactions \geinto $\nu W Z$ and \geinto $eW^+W^-$),
the two-standard deviation errors on \vtb \,
are shown in Tab. \ref{tab:7}.
\begin{table*}[htbp]\centering
\caption[ ]{Two-standard deviation error of \vtb \, expected for the
  annual luminosities as indicated. }
\begin{tabular}{lllll}    
                                &      &      &      &       \\
\hline\noalign{\smallskip}
                                &      &      &      &       \\
\SSQRTSEE, TeV                    &  0.5 & 1.0  & 1.5  & 2.0   \\
                                &      &      &      &       \\
\hline\noalign{\smallskip}
\hspace{3cm} & \hspace{2cm} & \hspace{2cm} & \hspace{2cm} & \hspace{2cm} \\
 $\cal{L}$ fb$^{-1}$        &  50  & 200 & 300   & 500   \\
 & &  &  & \\
 $\delta$\vtb        &  8\% & 2\% & 1.5\% & 1\% \\
 & &  &  & \\
\hline\noalign{\smallskip}
\end{tabular}
\label{tab:7}
\end{table*}
As can be seen, the CKM matrix element \vtb \, can be probed with high
accuracy.
Since $\delta$\vtb \, is proportional to 1/$\sqrt{N}$,
where $N$ is the number of events expected,
the \vtb \, measurement accuracy anticipated from reaction
(\ref{eq:2}) is competitive also at lower luminosities.
We would like to emphasize that the accuracy of \vtb \, expected at
\SSQRTSEE $\simeq$ 0.5 TeV is very similar to the expectations from the
Tevetron and the LHC \cite{heinson}. 
In particular, Willenbrock cited \cite{willenbrock} the following
one-standard deviation errors for \vtb: $\sim$10\% for Tevatron (Run
II), $\sim$5\% for Tevatron (Run III) and $\sim$5\% for the LHC.
However, an \ee \, collider at energies \SSQRTSEE $\gess$ 1.5 TeV
provides by means of the reaction \geinto $\nu b t$ a somewhat better
determination of \vtb \, even for half of the luminosities anticipated
in Tab. \ref{tab:7}.

\section{Probing the \hww coupling}
\label{sec6}

Reaction (\ref{eq:main}), \geinto \nbbw, involves also 
significant Higgs production   
(according to the diagrams (1)-(3) in Fig. \ref{fig:feyn}) 
with a rate directly proportional to the \hww \, coupling.
In recent years, the SM of electroweak interactions has
been beautifully confirmed.
In particular, $Z$ production and its decay at LEP-1 provided
$Z$-two-fermion couplings to be in agreement with theoretical predictions
at a 1\% level or better.
The bosonic sector however has been much
less investigated, mainly due to low energies available up to now.
With the onset of next generation of \ee \, linear colliders studies
of gauge boson interactions with e.g. the Higgs boson become crucial
in order to
find out how the SU(2)$\otimes$U(1) symmetry is broken.
In the SM the Higgs-vector boson vertices are uniquely determined.
Deviations from these couplings can occur in models with e.g. non-pointlike
character of the bosons or through interactions beyond the SM at high
energy scales.
In the following we do not specify a particular model to search for
non-SM coupling effects; rather we consider a class of models
which 
can be parametrized
by introducing an effective non-renormalizable Lagrangian which
preserves the SM gauge group
\begin{equation}
{\cal L}_{eff} = {\cal L}_{SM} +
\sum_{k=1}^{\infty}\frac{1}{(\Lambda^2)^k} \sum_i
f_i^{(k)}Q_i^{d_k} \quad ,
\label{eq:lagrangian}
\end{equation}
where $d_k = 2 k + 4$ denotes the dimension of operators and
$\Lambda$ is the energy scale of new interactions.
We limit ourselfs to 
the complete set of the effective dimension-6 operators as 
outlined in ref. \cite{buchmueller}.
Under this restriction phenomenological
applications to anomalous Higgs couplings have been discussed in
\cite{hagiwara1}-\cite{kilian}. 
In our study we adopted the notation of ref. \cite{hagiwara2} in which the
effective Lagrangian contains only four operators to
describe the \hww coupling.
Further limitation to a custodial SU(2) symmetry \cite{custodial}
as proposed in ref. \cite{kilian}, leads to only
two operators relevant for the Higgs process
(\ref{eq:3}), \geinto $\nu W H$:
\begin{equation}
\frac{1}{\Lambda^2}\left\{\frac{1}{2}f_{\varphi}\partial_{\mu}
(\Phi^+\Phi)\partial^{\mu}(\Phi^+\Phi)
+ f_{WW}\Phi^+({\hat W}_{\mu\nu}\hat W^{\mu\nu})\Phi\right\}.
\label{eq:1overlambda}
\end{equation}
Introducing such an effective \hww interaction in the program package
\CO \, and varying the parameters $F_i$ defined by
\begin{eqnarray}
F_{\varphi}/(1TeV^2) & = & f_{\varphi}/\Lambda^2 \hspace{1cm}
\qquad {\rm and} \nonumber\\ 
F_{WW}/(1TeV^2)& = & f_{WW}/\Lambda^2
\label{eq:F}
\end{eqnarray}
within 'reasonable' ranges, the impact on the cross section of
reaction (\ref{eq:3}) is investigated.
Natural values of $F_i$ are of $\cal O$(1);  when anomalous
contributions to the \hww coupling vanish, i.e. $F_i$ \into 0, the
SM is recovered.
For simplicity, the reaction (\ref{eq:3})
 cross sections are now calculated in the
unitary gauge instead of the  t'Hooft-Feynman gauge used so far
in the paper.
We have also checked that they are gauge invariant within the accuracy
achieved. 
The corresponding Feynman rules which follow from the Lagrangian
(\ref{eq:lagrangian}) and (\ref{eq:1overlambda}) 
are presented in the Appendix. 

\begin{figure}[htbp]
\mbox{\epsfxsize=17cm\epsfysize=9cm\epsffile{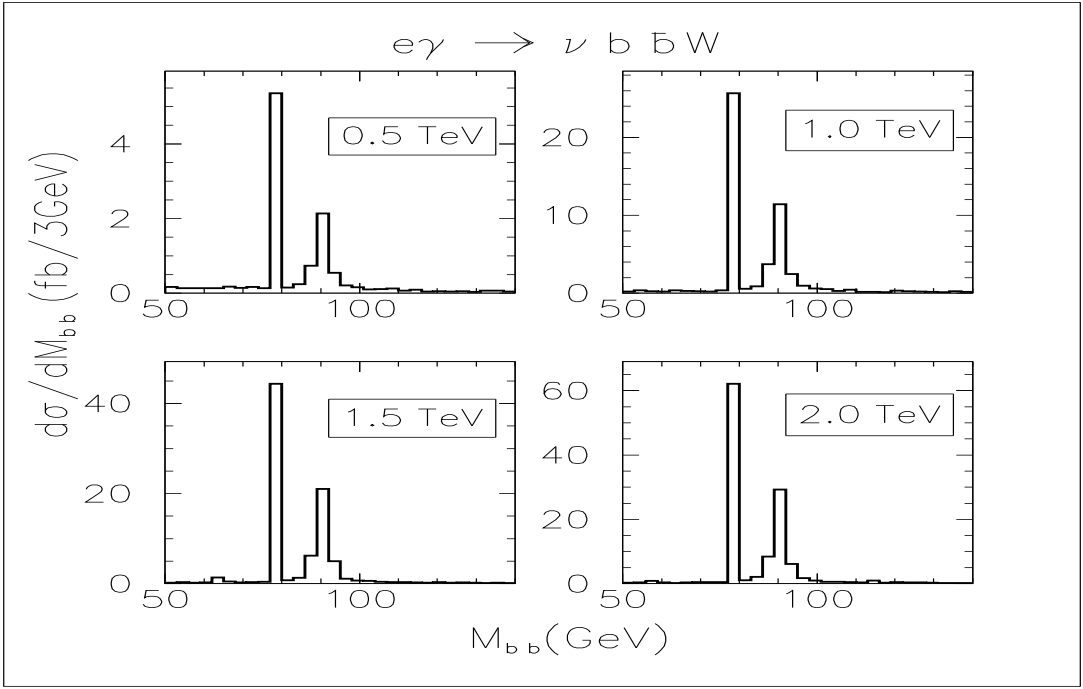}}
\caption{Differential cross sections as a function of the \bb \,
  invariant mass at \ee \, cm energies of 0.5, 1.0, 1.5 and 2.0 TeV. 
  Clear \hnull
  \, and $Z$ peaks are visible on a very small background.}
\label{fig:fmassb}
\end{figure}

\begin{figure}[htbp]
\mbox{\epsfxsize=17cm\epsfysize=9cm\epsffile{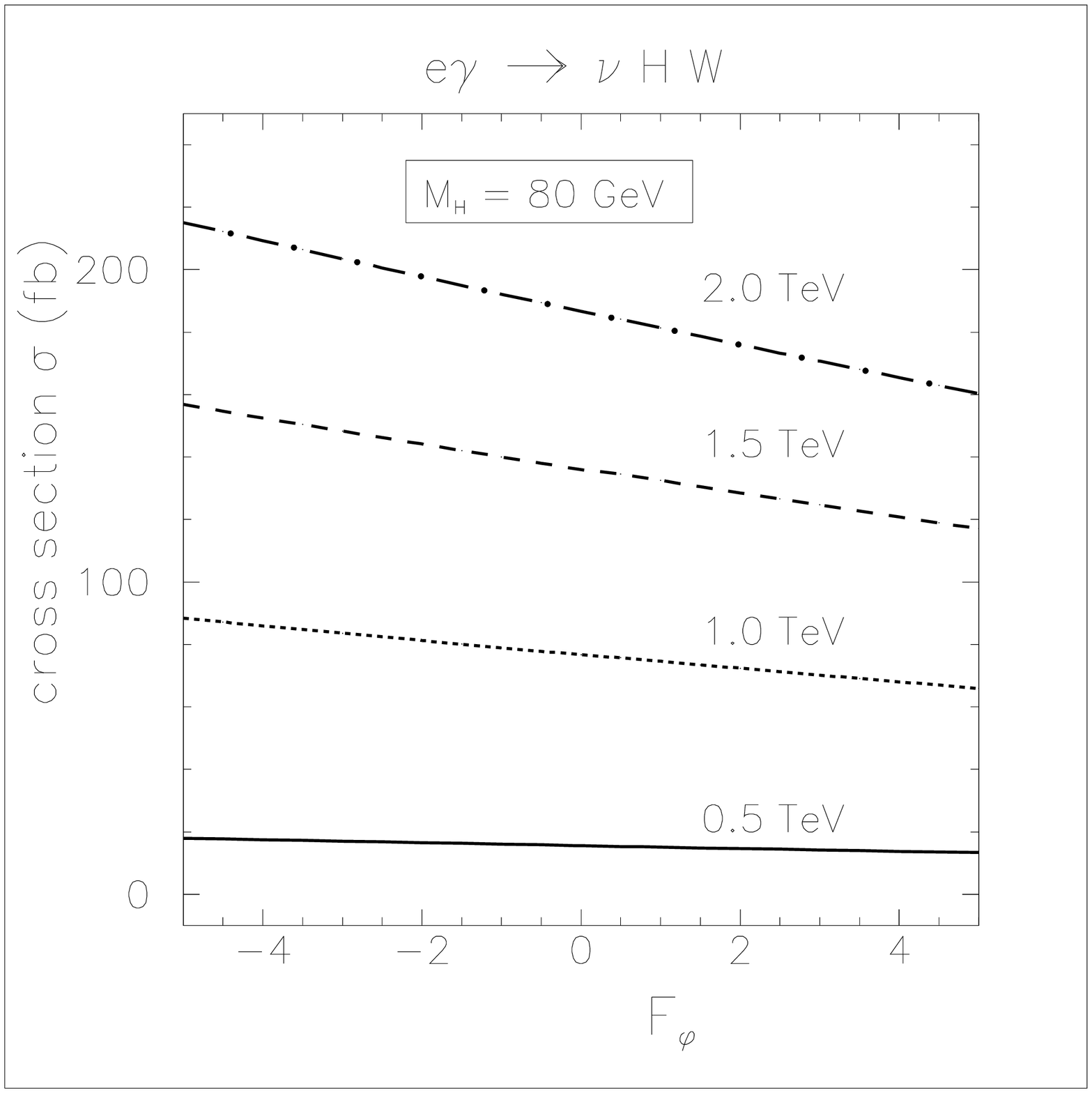}}
\caption{Cross sections for reaction (3)
as function 
of the parameter $F_{\varphi}$ with
  $F_{WW}$ = 0, at \ee \, cm energies  of  
  0.5, 1.0, 1.5 and 2.0 TeV for \mh =
  80 GeV.}
\label{fig:5}
\end{figure}

\begin{figure}[htbp]
\mbox{\epsfxsize=17cm\epsfysize=9cm\epsffile{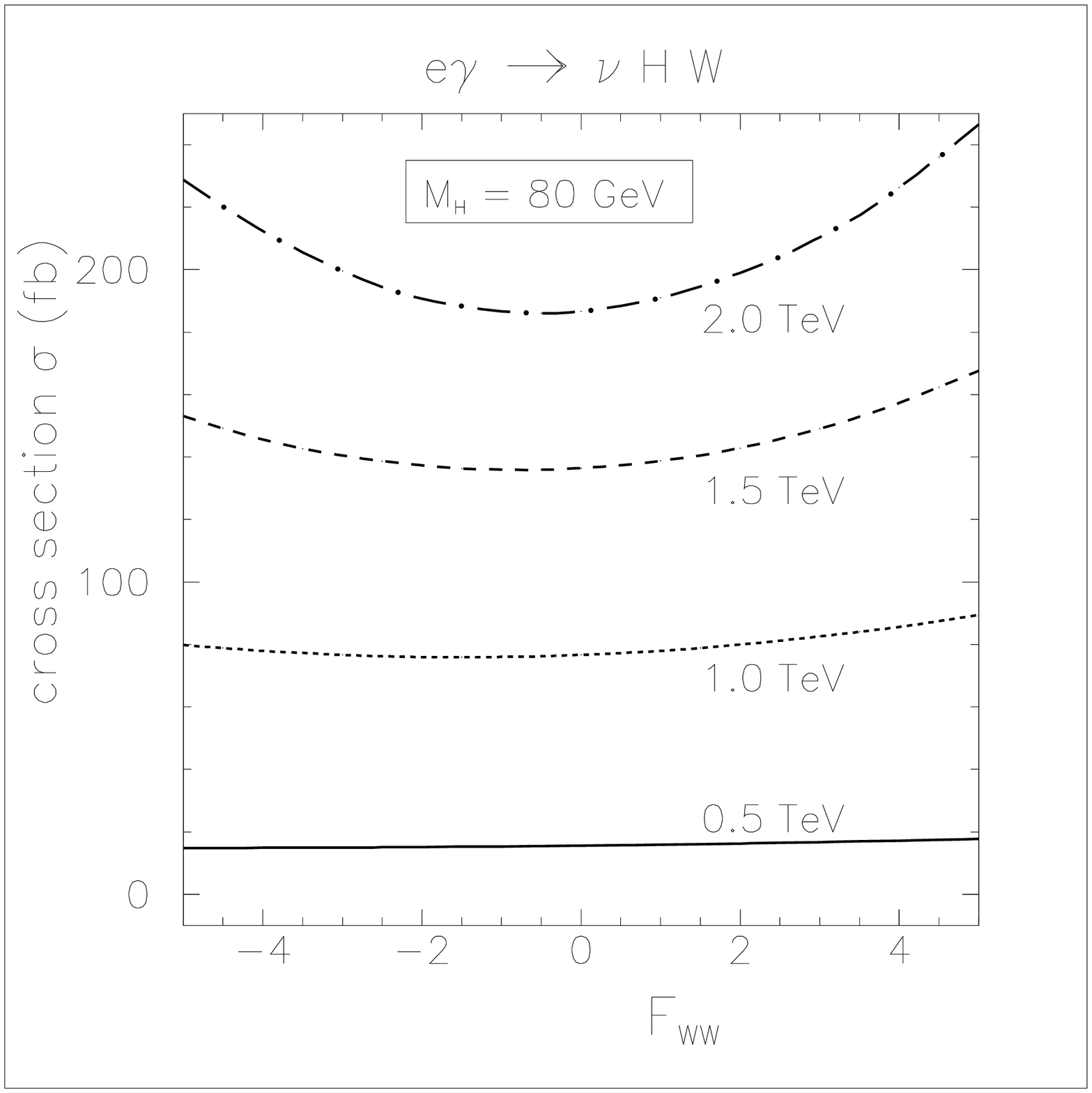}}
\caption{Cross sections for reaction (3) as functions 
of the parameter $F_{WW}$ with
  $F_{\varphi}$ = 0, at \ee \, cm energies of 0.5, 1.0, 1.5 and 2.0 TeV for \mh
  = 80 GeV.}
\label{fig:6}
\end{figure}

Probing the \hww \, coupling involves calculating the 
dependence of the cross section of reaction (\ref{eq:3}) on the 
parameters $F_i$ and comparison with the SM expectation.
Because of the restriction to the dimension-6 operators from the
beginning, the parameters $F_i$ cannot be varied simultaneously and the
cross section should only depend linearly on $F_i$. Otherwise,
  nonlinear terms in $F_i$ or mixed terms of $F_iF_j$ with an energy
  scale dependence of $\Lambda^{-4}$ would occur. 
Such terms are however
  excluded from our study by omitting higher than dimension-6
  operators.
The $\nu W H$ events are easily extracted from the 4-body final state $\nu b
\bar{b}W$ of reaction (\ref{eq:main}) by imposing a cut on the
\bb \, invariant mass, \mh - 3 GeV $< M(b\bar{b}) <$ \mh + 3 GeV, 
as seen in Fig. \ref{fig:fmassb}. 
Fig. \ref{fig:5}(\ref{fig:6}) shows the Higgs cross section as
function of $F_{\varphi}(F_{WW})$ for $F_{WW}(F_{\varphi})$ = 0, at \SSQRTSEE
= 0.5, 1.0, 1.5 and 2.0 TeV for \mh = 80 GeV.
In both Figures, the parameter $F_i$ ranges between -5 and +5 and, as
can be seen, deviations from linearity occur at higher energies.
In order to estimate the ranges of $F_i$ which can
be probed within our assumptions, we determined those variations of
the $F_i$ which leave the cross section unchanged within 2 s.d. 
from the SM value.
Only statistical errors of the cross sections were
considered taking into account the integrated luminosities of
Tab. \ref{tab:7}, an $e-\gamma$ conversion factor of 0.8 and a 30\%
$\nu H W$ detection probability.
In addition, it has been checked that the cross sections in the $F_i$
intervals  so obtained are in accord with a linear behaviour
as required by the restriction on dimension-6 operators.
The intervals of $F_i$ obtained are presented in
Tab. \ref{tab:8}.
\begin{table*}[htbp]\centering
\caption[ ]{Range of   $|F_{\varphi}|$ and  $|F_{WW}|$ obtained from the
  two-standard deviation criteria as described
  in the text.}

\begin{tabular}{lllll}    
                                &      &      &      &       \\
\hline\noalign{\smallskip}
                                &      &      &      &       \\
\SSQRTSEE, TeV                    &  0.5 & 1.0  & 1.5  & 2.0   \\
                                &      &      &      &       \\
\hline\noalign{\smallskip}
\hspace{3cm} & \hspace{2cm} & \hspace{2cm} & \hspace{2cm} & \hspace{2cm} \\
 $|F_{\varphi}|$                   &  5.0  & 1.0  & 0.6   & 0.4   \\
                                &      &     &       &       \\
 $|F_{WW}|$                     &  9.0  & 2.5 & 2.0    & 1.0    \\
                                &      &     &       &       \\
\hline\noalign{\smallskip}
\end{tabular}
\label{tab:8}
\end{table*}

Clearly, only at energies $\gess$ 1 TeV the total cross section of
\geinto $\nu W H$ involves some sensitivities to the anomalous
couplings considered.
Recent analyses of e.g. the simpler two-body reactions \eeinto $H
\gamma$ or $H Z$ \cite{gounaris,kilian} revealed high sensitivities for these
operators including production and decay angular distributions.
We expect that by inclusion of differential distributions or the whole
phase space event population the sensitivity with respect to the
$|F_i|$ becomes larger.
Such a study including additional 3-body final states is in
preparation and will be published in a forthcoming paper.
At present, we would like to point out that the process \geinto $\nu W
H$ is of interest at \SSQRTSEE = 1-2 TeV since it becomes sensitive and
comparable to other reactions to probe the structure of operators.
In this respect the disentangling of the origin of 'new physics' may
be further helped by comparing analyses of different reactions. 
\section{Summary}
\label{sec7}

Results of a complete tree-level calculation for the reaction \geinto
\nbbw \, at cm energies  0.5 to 2.0 TeV are presented and discussed,
using the computer package \CO.
This reaction is very interesting on its own because it involves 
at the same time single top production, \geinto $\nu \bar{b} t$, and
associated Higgs production, \geinto $\nu H W$, with subsequent decays
of $t$ \into $W b$ and $H$ \into \bb, respectively. 
Therefore, both three-body reactions already studied in previous
publications are analyzed in an extended manner taking into account
interferences between different subchannels and the
irreducible background.

We present the total cross sections of reaction (\ref{eq:main}) as well
as those of its 
main components as functions of \SSQRTSGE \, and \SSQRTSEE, the
latter after convolution with the backscattered photon flux of
ref. \cite{ginzburg}.
Above threshold, Higgs and $Z/\gamma^*$ (with  $H$ \into \bb \, and 
$Z/\gamma^*$
\into \bb \, decays) contribute with about equal weights to the total
rate while single top production is roughly a factor 2 lower.
The Higgs and the top production cross sections are only weakly
dependent on the Higgs mass in the range 
80 to 140 GeV and the top mass between
160 and 200 GeV.
The contributions from multiperipheral diagrams are very
small.
They grow however with increasing energy.
Interferences between different subchannels 
were found to be significant only at the highest energies.
They are
to some extent compensated by the multiperipheral contributions.

The event rate for the reaction \geinto $\nu t b$, which is large even after 
folding with an energy spectrum of the backscattered photon beam
and making reasonable assumptions on collider 
luminosities and detection probabilities, 
provides a very sensitive measurement for the CKM
matrix element \vtb.
If the cross section for single top production is measured 
with high accuracy, the two-standard deviation errors on \vtb \, can be
close to (1-3)\% at \SSQRTSEE = 1-2 TeV.
To our knowledge such an accuracy cannot be achieved by other
measurements so far considered.

The reaction \geinto $\nu H W$ allows to probe the \hww \, coupling
and to measure parameters of dimension-6 operators in the effective Lagrangian.
It has been found that at 0.5 TeV the accuracy obtained on these
parameters is not sufficient to make this measurement 
sensitive to new physics while at energies \SSQRTSEE = 1-2 TeV
the \hww \, coupling can be probed with high sensitivity and
deviations from the Standard Model could show up.

\appendix{Appendix}

The Feynman rules for \hww \, and  \hww$\gamma$
vertices in unitary gauge which follow from the effective Lagrangian
(\ref{eq:lagrangian}) and (\ref{eq:1overlambda}): 

\begin{eqnarray}
\Gamma^{HWW}_{\mu \nu}(p,q,\kappa) = \frac{e M_W}{s_W}\biggl\{(1 -
\frac{1}{4}f_{\varphi}\frac{v^2}{\Lambda^2})g_{\mu\nu} \quad & + & 
\nonumber\\  [1cm] +  \quad 
2 f_{WW}\frac{1}{\Lambda^2}[g_{\mu\nu}(q,\kappa) - 
q_{\nu}\kappa_{\mu})]\biggr\} & &
\end{eqnarray}

\noi and 

\begin{eqnarray}
\Gamma^{HWW}_{\mu \nu\alpha}(p,q,\kappa, l) = \frac{e^2 M_W}{s_W} 2
f_{WW} \frac{1}{\Lambda^2}
& \biggl\{ & g_{\mu\nu}(q - \kappa)_{\alpha} \quad  -  \nonumber\\
 -  \quad q_{\nu}g_{\mu\alpha} + \kappa_{\mu}g_{\nu\alpha}\biggr\} & &
\end{eqnarray}

where $v = \frac{2 M_W}{e}s_W$ is the vacuum expectation value;
$p,q,\kappa$ and $l$ are the momenta of the $H, W^+, W^-$ and
$\gamma$ fields, 
respectively.
The Lorentz indices of the  $W$'s and $ \gamma$
fields  are denoted as $\mu, \nu$ and $ \alpha$, respectively.
The quantity 1 in the first term of $\Gamma^{HWW}_{\mu \nu}$ corresponds to the
SM vertex.
The second vertex $\Gamma^{HWW}_{\mu \nu\alpha}$ does not occur in the SM
at tree level.

\section*{Acknowledgements}
E.B. and A.P. are grateful to DESY IfH Zeu\-then for the kind hospitality,
and  to P. S\"oding for his interest and support.
The work has been supported in part by the 
RFBR grants 96-02-19773a and 96-02-18635a, and by the grant 95-0-6.4-38 of 
the Center for Natural Sciences of State Committee for Higher Education
in Russia. 




\end{document}